\documentclass[a4paper,11pt,aps,preprintnumbers,amsmath,amssymb,superscriptaddress,nofootinbib]{article}
\pdfoutput=1
\usepackage{jheppub}

\usepackage{amssymb}
\usepackage{enumerate}
\usepackage{amssymb}
\usepackage{amsmath}
\usepackage{tikz}
\usepackage{feynmf}
\usepackage{makecell}
\usepackage{slashed}
\usepackage{natbib}
\usepackage{graphicx}
\hypersetup{colorlinks=true}

\usepackage{mathrsfs,amssymb}  
\usepackage{cancel}
\usepackage[normalem]{ulem}

\definecolor{darkgreen}{rgb}{0.01, 0.75, 0.24}
\definecolor{darkorange}{rgb}{1.0, 0.55, 0.0}
\usepackage{pifont}
\newcommand{\cmark}{{\color{darkgreen} \ding{52}}}%
\newcommand{\comark}{{\color{darkorange} \ding{51}}}%
\newcommand{\xmark}{{\color{red} \ding{55}}}%

\newcommand\be{\begin{equation}}
\newcommand\ee{\end{equation}}
\newcommand{\comment}[1]{}
\newcommand\bea{\begin{eqnarray}}
\newcommand\eea{\end{eqnarray}}

\preprint{
\begin{minipage}{3cm}
\small
\flushright
MI-TH-2019
\end{minipage}}

\title{Implications of the XENON1T Excess on the Dark Matter Interpretation}

\author{Haider Alhazmi$^{1,2}$,} 
\author{Doojin Kim$^3$,}
\author{Kyoungchul Kong$^1$,}
\author{Gopolang Mohlabeng$^4$,}
\author{Jong-Chul Park$^5$}
\author{and Seodong Shin$^6$}
\affiliation{
$^1$Department of Physics and Astronomy, University of Kansas, Lawrence, KS 66045, USA \\
$^2$Department of Physics, Jazan University, Jazan 45142, Saudi Arabia\\
$^3$Mitchell Institute for Fundamental Physics and Astronomy, Department of Physics and Astronomy, Texas A\&M University, College Station, TX 77845, USA\\
$^4$ Physics Department, Brookhaven National Laboratory, Upton, New York 11973, USA\\
$^5$ Department of Physics and Institute of Quantum Systems (IQS), Chungnam National University, Daejeon 34134, Korea\\
$^6$ Department of Physics, Jeonbuk National University, Jeonju, Jeonbuk 54896, Korea\\ 
}
\emailAdd{haider@ku.edu}
\emailAdd{doojin.kim@tamu.edu}
\emailAdd{kckong@ku.edu}
\emailAdd{gmohlabeng@bnl.gov}
\emailAdd{jcpark@cnu.ac.kr}
\emailAdd{sshin@jbnu.ac.kr}

\abstract{
The dark matter interpretation for a recent observation of excessive electron recoil events at the XENON1T detector seems challenging because its velocity is not large enough to give rise to recoiling electrons of $\mathcal{O}({\rm keV})$. 
Fast-moving or boosted dark matter scenarios are receiving attention as a remedy for this issue, rendering the dark matter interpretation a possibility to explain the anomaly.
We investigate various scenarios where such dark matter of spin 0 and 1/2 interacts with electrons via an exchange of vector, pseudo-scalar, or scalar mediators. 
We find parameter values not only to reproduce the excess but to be consistent with existing bounds. 
Our study suggests that the scales of mass and coupling parameters preferred by the excess can be mostly affected by the type of mediator, and that significantly boosted dark matter can explain the excess depending on the mediator type and its mass choice.
The method proposed in this work is general, and hence readily applicable to the interpretation of observed data in the dark matter direct detection experiment. 
}

\begin{document}

\maketitle

\section{Introduction}

Dark matter is a crucial ingredient in the cosmological history of the universe and accounts for about 27\% of the energy budget in the universe today. 
As its existence is supported by galactic-scale to cosmological-scale gravity-based evidence, various experiments were performed, are operational, and are planned to detect dark matter via its hypothetical non-gravitational interactions with Standard Model (SM) particles.
While no conclusive observations have been made thus far, the XENON Collaboration has recently reported an excess of electron recoil events over known backgrounds with an exposure of 0.65~ton$\cdot$year~\cite{Aprile:2020tmw}.
The excess is shown below 7~keV and most of the events populate at $2-3$~keV.

The XENON1T detector is designed to have an extremely low rate of background events, so this excess could be considered as a sign of new physics.
The XENON Collaboration has claimed that while the unresolved $\beta$ decays of tritium can explain the excess at 3.2$\sigma$ significance, the solar axion model and the neutrino magnetic moment signal can be favored at 3.5$\sigma$ and 3.2$\sigma$ significance, respectively.
It is expected that confirmation or rejection of these hypotheses will be done with more statistics in the near future.
By contrast, the interpretation with conventional dark matter is less favored, essentially because of its non-relativistic nature.
For dark matter sufficiently heavier than electron, the scale of electron recoil (kinetic) energy is $\sim m_e \times(10^{-3}c)^2 \approx \mathcal{O}({\rm eV})$ with $m_e$ and $10^{-3}c$ being the mass of electron and the typical velocity of dark matter near the earth, respectively.
In other words, the energy deposition by conventional dark matter is not large enough to accommodate the excessive events of $\mathcal{O}({\rm keV})$.

However, this issue may be avoided by envisioning non-conventional dark-sector scenarios involving a mechanism to exert a sufficient boost on a dark matter component, rendering the dark matter hypothesis plausible enough to explain the excess.
In particular, upon confirmation, the XENON1T anomaly can be the first signal to indicate that the associated dark sector is non-conventional, opening a new pathway toward dark matter phenomenology.
Indeed, the authors in Ref.~\cite{Giudice:2017zke} pointed out, for the first time, that the XENON1T experiment would be sensitive enough to the fast-moving $\chi_1$ -- which arises in the two-component boosted dark matter (BDM) scenario -- interacting with electrons.
Along this line, we entertain a class of non-conventional dark-sector scenarios to explain the XENON1T excess in this paper, in particular, focusing on the impact of the particle mediating the dark-matter--electron interactions in the context of the BDM scenario as a concrete example.

\section{Dark Matter Interpretation}

As mentioned previously, it is challenging to accommodate the XENON1T anomaly using the ordinary halo dark matter since its typical velocity is too small to invoke keV-scale energy deposition on target electrons.
Bosonic dark matter (e.g., axion-like particle and dark photon) of keV-scale mass could be absorbed, depositing its whole mass energy in the XENON1T detector.
However, this is likely to give rise to a line-like signature, so that this possibility is less preferred by the observed recoil energy spectrum.
Indeed, the XENON Collaboration found that no bosonic dark matter of mass within 1 and 210~keV shows more than 3$\sigma$ significance unlike the other interpretations, so they simply set the limits for relevant dark matter candidates.

The upshot of this series of observations is that dark matter (or more generally, a dark matter component) should acquire a sizable enough velocity to transfer keV-scale kinetic energy to a target electron.
This approach has been investigated in Ref.~\cite{Kannike:2020agf} where the authors claimed that fast-moving dark matter with velocity of $\mathcal{O}(0.1 c)$ can fit in the XENON1T excess.
An important implication of this way of dark matter interpretation is that the dark matter (candidate) responsible for the excess is not the (cold) galactic halo dark matter, i.e., it is a subdominant fast-moving component and the underlying dark matter scenario is non-conventional.~\footnote{See also Refs.~\cite{Su:2020zny, Bally:2020yid, Paz:2020pbc, Primulando:2020rdk, Jho:2020sku, An:2020tcg, Ge:2020jfn, Bhattacherjee:2020qmv} for other explanations of the excess.}
Furthermore, it requires a certain mechanism to ``boost'' this dark matter component in the universe today.
There are several mechanisms and scenarios to serve this purpose, which were originally proposed for other motivations; semi-annihilation~\cite{DEramo:2010keq}, (two-component) boosted dark matter scenarios~\cite{Belanger:2011ww,Agashe:2014yua,Kim:2016zjx}, models involving dark-matter-induced nucleon decays inside the sun~\cite{Huang:2013xfa}, and energetic cosmic-ray-induced dark matter~\cite{Yin:2018yjn, Bringmann:2018cvk, Ema:2018bih}. 
Of them, we discuss the BDM scenario as also considered in Ref.~\cite{Fornal:2020npv}, focusing on implications of the XENON1T anomaly on the spins of BDM and mediator particles.

The standard two-component BDM scenario~\cite{Agashe:2014yua} assumes two different dark matter species; one (say, $\chi_0$) is heavier than the other (say, $\chi_1$).
Their stability is often protected by separate unbroken symmetries such as $Z_2\otimes Z_2'$ and ${\rm U}(1)'\otimes{\rm U}(1)''$.
One of the species (usually the heavier one $\chi_0$) has no direct coupling to SM particles, but communicates with the other species $\chi_1$.
By contrast, $\chi_1$ can interact with SM particles with a sizable coupling.
Therefore, $\chi_0$ is frozen out via the indirect communication with the SM sector with the ``assistance'' of $\chi_1$ (a.k.a. ``assisted'' freeze-out mechanism)~\cite{Belanger:2011ww}.
In other words, $\chi_0$ pair-annihilates to $\chi_1$ while $\chi_1$ pair-annihilates to SM particles.
The relatively sizable coupling of $\chi_1$ to SM particles renders it the negligible dark matter component while keeping $\chi_0$ dominant in the galactic halo.
In most of the well-motivated parameter space, conventional dark matter direct detection experiments do not possess meaningful sensitivity to relic $\chi_0$ and $\chi_1$ because of tiny coupling and negligible statistics, respectively. 

A phenomenologically intriguing implication of this model setup, particularly relevant to the XENON1T excess, is that $\chi_1$ can acquire a sizable boost factor, which is simply given by the ratio of the $\chi_1$ mass to the $\chi_0$ mass, in the universe today.
Therefore, one may look for the signal induced by such boosted $\chi_1$.
Due to the small $\chi_1$ flux (see also Eq.~\eqref{eq:flux}), it is usually challenging for small-volume detectors to have signal sensitivity, but ton-scale dark matter direct detection experiments can be sensitive to the boosted $\chi_1$ signal~\cite{Cherry:2015oca,Giudice:2017zke}. 
As mentioned earlier, Ref.~\cite{Giudice:2017zke} has performed the first sensitivity study for the boosted $\chi_1$ interacting with electrons in XENON1T, LUX-ZEPLIN, and DEAP3600 experiments.
Motivated by the proposal in Ref.~\cite{Giudice:2017zke}, the COSINE-100 Collaboration has conducted the first search for BDM-induced signals as a dark matter direct detector\footnote{Note that Super-Kamiokande, a 10 kton-scale neutrino detector, performed a dedicated search for BDM interacting with electrons~\cite{Kachulis:2017nci}.} and reported the results~\cite{Ha:2018obm} including limits on the models of inelastic BDM~\cite{Kim:2016zjx}.

Denoting the $\chi_0$ and $\chi_1$ mass parameters by $m_0$ and $m_1$ correspondingly, we find that if $m_1$ is given by approximately $99.0 - 99.9\%$ of $m_0$, $\chi_1$ coming from the pair-annihilation of $\chi_0$ in the present universe can be as fast-moving as $0.04-0.14c$.
While this simple consideration determines the ``desired'' mass relation between $\chi_0$ and $\chi_1$, not all mass values are favored by the excess aside from the various existing limits.
More importantly, as will be discussed later, the ``favored'' velocity range can be significantly altered, depending on the underlying mass spectrum and particle types.
To investigate these points more systematically, we first consider the number of signal events $N_{\rm sig}$.
As well known, it is given by
\begin{equation}
    N_{\rm sig} = \mathcal{F}_1\,\sigma_{1e}\, N_{e,\,{\rm tot}}^{\rm eff} \,t_{\rm exp}\,, \label{eq:Nsig}
\end{equation}
where $\mathcal{F}_1$, $\sigma_{1e}$, $N_{e,\,{\rm tot}}^{\rm eff}$, and $t_{\rm exp}$ are the flux of boosted $\chi_1$ near the earth, the total scattering cross-section of $\chi_1$ with an electron, the number of effective target electrons in the fiducial volume of the XENON1T detector, and the total exposure time, respectively.
Here $\sigma_{1e}$ could be affected by the threshold and/or detection efficiencies for recoiling electrons if a significant number of events are populated in the region where the recoil electron energy is near the threshold and/or the associated efficiencies are not large enough.
The last two factors are experimentally determined and their product can be easily deduced from 0.65 ton$\cdot$year.

Regarding $N_{e,\,{\rm tot}}^{\rm eff}$, we remark that the binding energy of electrons in the xenon atom is not negligible given the keV scale of recoiling electron kinetic energy.
While the outermost electron (in the $O$ shell) needs 12.1~eV~\cite{1978ps1..book.....C} to get ionized, the innermost electron (in the $K$ shell) requires an ionization energy of 34.6~keV~\cite{Bearden:1967gqa}.
Therefore, only some fraction of electrons can be target electrons for the BDM mostly inducing keV-scale energy deposition.
Some works considered form factors to calculate the dark matter event rate to explain the XENON1T excess.
For example, Ref.~\cite{Kannike:2020agf} used the atomic excitation factor with relativistic corrections and Ref.~\cite{Cao:2020bwd} considered the dark matter and ionization form factors, restricting to the $N$-shell and $O$-shell electrons. We here take a shortcut scheme, reserving a dedicated analysis for future work~\cite{AKKMPS}. 
As a conservative approach, we consider electrons from three outermost orbitals ($5p$,$5s$ and $4d$), which are known to be the dominant contribution~\cite{Essig:2011nj,Lee:2015qva,Cao:2020bwd}, i.e., the number of target electrons in a single xenon atom $N_e^{\rm eff}$ is taken to be 18 throughout our analysis.\footnote{For 1 ton of liquid xenon, $N_{e,{\rm tot}}^{\rm eff} = 4.59 \times 10^{27} \, N_e^{\rm eff}$.}
Note that the largest ionization energy among the electrons belonging to the three orbitals is $\sim76$~eV~\cite{1978ps1..book.....C} which would induce less than 5\% uncertainty in estimating $2-3$~keV energy deposition.
Since we will consider energy resolution of $\sim 450$~eV~\cite{Aprile:2020yad}, 
we expect that the $\lesssim0.1$~keV level uncertainty is buried in the detector resolution.
We also note that each of the $N$-shell and $O$-shell electrons gets excited with a different weight.
We expect that this would make an $\mathcal{O}(1)$ effect, so our findings and conclusions in the analysis would remain valid. 
We will revisit this aspect before we conclude our study.

\begin{figure}[t]
    \centering
    \includegraphics[width=11cm]{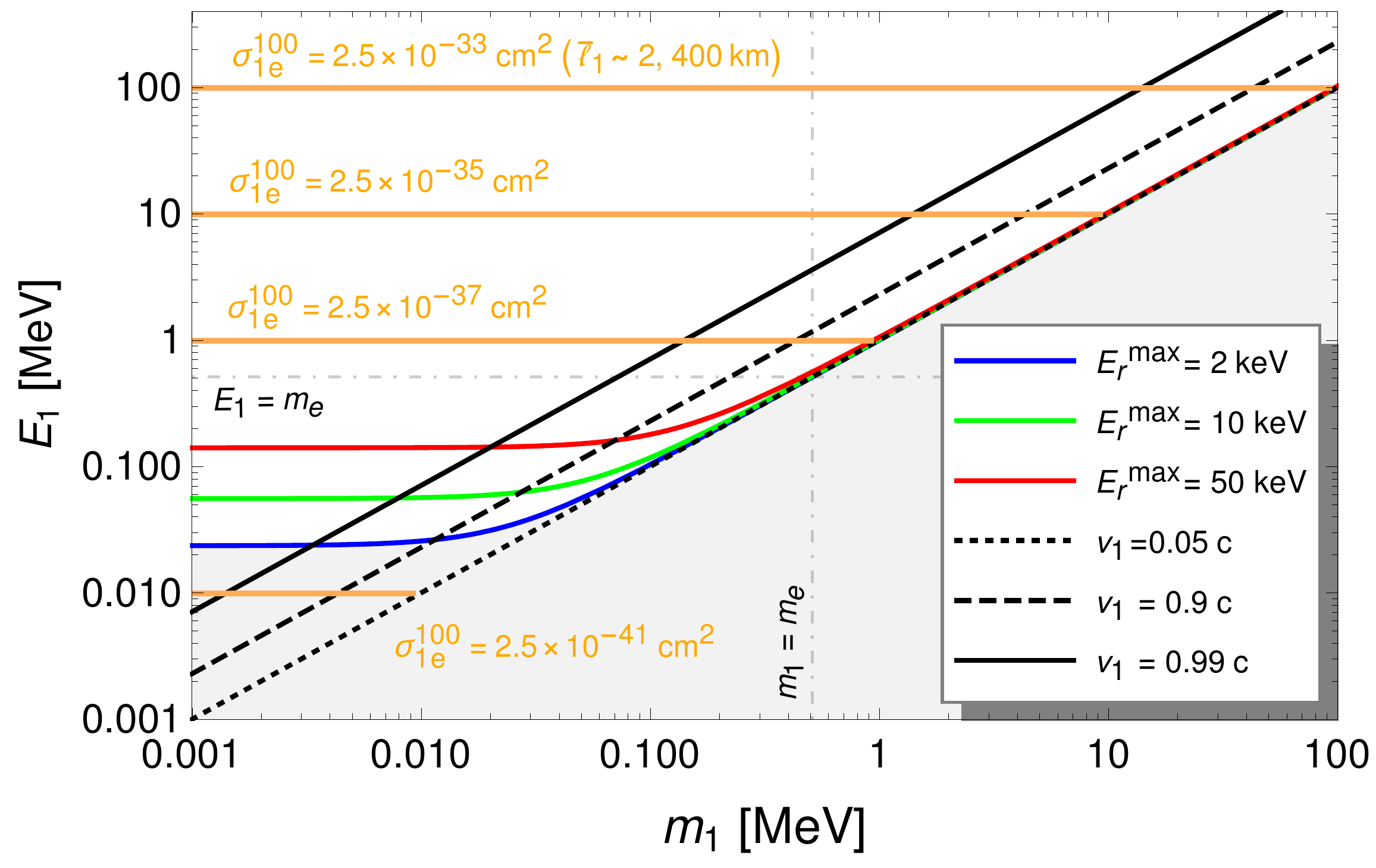}
    \caption{Maximum recoil energy of electrons scattered off by BDM, $E_r^{\rm max}$ (solid colored curves), and required BDM-electron scattering cross-sections to have 100 recoil events with the 0.65 ton$\cdot$year exposure, $\sigma_{1e}^{100}$ (orange lines), in the $(m_1, E_1)$ plane.
    The gray-shaded lower-right area is disfavored because the expected maximum electron recoil energy is less than the typical energy associated with the observed excess, i.e., $E_r^{\rm max} < 2$ keV.
    The upper region requires large cross-sections which can result in too small mean free paths ($\bar{\ell}_1 \propto 1/\sigma_{1e}$) inside the earth to reach the XENON1T detector.
    We show a mean free path value at $E_1=100$~MeV for reference.
    Three diagonal lines represent the velocity of BDM for a given choice of the $(m_1, E_1)$ pairs.
    }
    \label{fig:m0-m1}
\end{figure}

The estimate of flux $\mathcal{F}_1$ depends on the source of BDM, and we consider here the $\chi_1$ coming from the galactic halo for illustration. 
Assuming that the $\chi_0$ halo profile follows the Navarro-Frenk-White profile~\cite{Navarro:1995iw,Navarro:1996gj}, we see that $\mathcal{F}_1$ from all sky is given by~\cite{Agashe:2014yua}
\begin{eqnarray}
    \mathcal{F}_1 = 1.6~{\rm cm}^{-2}{\rm s}^{-1}               \times  \left(\frac{ \langle \sigma_{0\to 1}v\rangle }{5\times 10^{-26}~{\rm cm}^3{\rm s}^{-1}} \right)\left(\frac{10~{\rm MeV}}{m_0} \right)^2\,, \label{eq:flux}
\end{eqnarray}
where the velocity-averaged annihilation cross-section $\langle \sigma_{0\to 1}v\rangle$ is normalized to $5\times 10^{-26}~{\rm cm}^3{\rm s}^{-1}$ to be consistent with the observed relic abundance.
Note that the flux is proportional to inverse mass square, so roughly speaking a large (small) $m_0$ prefers a large (small) value of $\sigma_{1e}$ to reproduce the excessive number of events of XENON1T.  

It is instructive to investigate the BDM parameter space to (potentially) accommodate the XENON1T anomaly in a model-independent fashion.
In figure~\ref{fig:m0-m1}, we present the maximum recoil energy of electrons scattered off by BDM,
\begin{equation}
E_r^{\rm max}=\frac{2m_ep_1^2}{s}\,, \label{eq:maxE} 
\end{equation}
where $p_1^2=E_1^2-m_1^2$ and $s=m_1^2+m_e^2+2m_eE_1$ with $E_1$ being the total energy of boosted $\chi_1$, and required BDM-electron scattering cross-sections to have 100 recoil events at the XENON1T detector with the 0.65 ton$\cdot$year exposure, $\sigma_{1e}^{100}$, in the $(m_1, E_1)$ plane.
Note that although the number of excessive events is about 50, the nominal number of signal events can be a few times larger due to detector efficiency and resolution, depending on the underlying model details.
In the two-component annihilating BDM scenario that we consider here, $E_1$ is simply identified as $m_0$.

$E_r^{\rm max}$ must be at least 2 keV because the observed excess is pronounced most at $2-3$ keV.
The disfavored region of $E_r^{\rm max} < 2$ keV is gray-shaded.
From Eqs.~(\ref{eq:Nsig}) and (\ref{eq:flux}), $N_{\rm sig} \propto \mathcal{F}_1 \sigma_{1e} \propto \sigma_{1e}/E_1^2$, so the required cross-section increases quadratically in $E_1$.
One should keep in mind that too large $\sigma_{1e}^{100}$ is constrained by too short a mean free path and (potentially) by various experimental bounds on the mediator mass and the associated coupling.
We will discuss these issues in the context of specific benchmark points later.

\begin{table}[t]
    \centering
    \resizebox{\columnwidth}{!}{
    \begin{tabular}{c|c|c|c|c}
    \hline \hline
    Case & Mediator & Dark matter & $\mathcal{L}_{\rm int}$ & $\overline{|\mathcal{A}|}^2$  \\
    \hline
    VF & $V_\mu$  & $\chi_1$ & $(g_e^V \bar{e} \gamma^\mu e + g_\chi^V \bar{\chi}_1\gamma^\mu\chi_1)V_\mu$ & $8m_e\left\{ m_e(2E_1^2-2E_1E_r+E_r^2)-(m_e^2+m_1^2)E_r\right\}$ \\
    VS & $V_\mu$ & $\varphi_1$ & $(g_e^V \bar{e}\gamma^\mu e+g_\varphi^V \varphi_1^*\partial^\mu \varphi_1+ {\rm h.c.})V_\mu$ & $8m_e\left\{ 2m_eE_1(E_1-E_r)-m_1^2E_r \right\}$ \\
    PF & $a$ & $\chi_1$ & $(ig_e^a\bar{e} \gamma^5 e+ig_\chi^a \bar{\chi}_1\gamma^5 \chi_1)a$ & $4m_e^2E_r^2$ \\
    PS & $a$ & $\varphi_1$ & $(ig_e^a\bar{e}\gamma^5 e+ ig_\varphi^a m_1\varphi^*\varphi )a$ & $8 m_e m_1^2E_r$ \\
    SF & $\phi$ & $\chi_1$ & $(g_e^\phi \bar{e}e+g_\chi^\phi \bar{\chi}_1 \chi_1)\phi$ & $4m_e(E_r+2m_e)(2m_1^2+m_eE_r)$ \\
    SS & $\phi$ & $\varphi_1$ & $(g_e^\phi \bar{e}e + g_\varphi^\phi m_1 \varphi^*\varphi)\phi$ & $8m_e m_1^2(E_r+2m_e)$ \\
    \hline \hline
    \end{tabular}
    }
    \caption{Example scenarios associated with the interaction between BDM and electron that we consider in this study.
    $V_\mu$, $a$, and $\phi$ denote vector, pseudo-scalar, scalar mediators, respectively, while $\chi_1$ and $\varphi_1$ denote (Dirac-)fermionic and (complex-)scalar dark matter.
    For the PS and SS cases, the scale of mediator couplings to dark matter is normalized to the mass of BDM for convenience.}
    \label{tab:models}
\end{table}

To study the model-dependence of the BDM scattering cross-section, we consider a vector mediator $V_\mu$, pseudo-scalar mediator $a$, and scalar mediator $\phi$ together with (Dirac-)fermionic BDM $\chi_1$ and (complex-)scalar BDM $\varphi_1$; the six different cases in total are summarized in Table~\ref{tab:models} with the relevant interaction terms and coupling constants.
For the PS and SS cases, the scale of mediator couplings to dark matter is normalized to $m_1$.
Assuming that the incoming $\chi_1$ is much faster than the electrons in xenon atoms, we find that the spectrum in the kinetic energy of recoiling electrons $E_r$ with incoming BDM energy $E_1$ has the form of
\begin{equation}
    \frac{d\sigma_{1e}}{dE_r}=\frac{(g_j^ig_e^i)^2m_e}{8\pi \lambda(s,m_e^2,m_1^2)(2m_eE_r+m_i^2)^2} \overline{|\mathcal{A}|}^2 \label{eq:xs} 
\end{equation}
where $i\in \{V,\,a,\,\phi\}$, $j\in \{\chi,\,\varphi\}$, and $\lambda(x,y,z)=(x-y-z)^2-4yz$.
Here $\overline{|\mathcal{A}|}^2$ is the spin-averaged amplitude squared, in which the denominator from the propagator contribution is factored out, and the expressions for the six cases are also tabulated in Table~\ref{tab:models}.

\section{Case Studies}

We are now in the position to look into the aforementioned cases, starting with ($a$) the vector mediator case, followed by ($b$) the pseudo-scalar mediator case and ($c$) the scalar mediator case.
Since the two dark matter components are assumed to be thermally produced, we assume that the mass of the heavier component (i.e., dominant relic) is larger than, at least, a few MeV.
To develop the intuition on this differential spectrum, we consider three different regions of mass space: 
\begin{eqnarray}
&(i)& m_0\approx m_1 \gg m_e,\quad m_i \gg m_e,\nonumber \\ 
&(ii)& m_0\approx m_1 \gg m_e,\quad m_i < m_e, \label{eq:conditions}\\ 
&(iii)& m_0 \gg m_e > m_1,\quad m_i < m_e\,,\nonumber
\label{eq:cases}
\end{eqnarray}
where $m_i$ is the mediator mass with $i=V,a,\phi$ and $m_0$ is again assumed to be greater than $m_1$ in all cases.
Note that $(i)$ and $(ii)$ represent the upper-right region of the $(m_1, E_1)$ parameter plane with respect to $(m_e, m_e)$ in figure~\ref{fig:m0-m1}, while $(iii)$ does the upper-left region. 

\medskip

\begin{figure}[t]
    \centering
    \includegraphics[width=11cm]{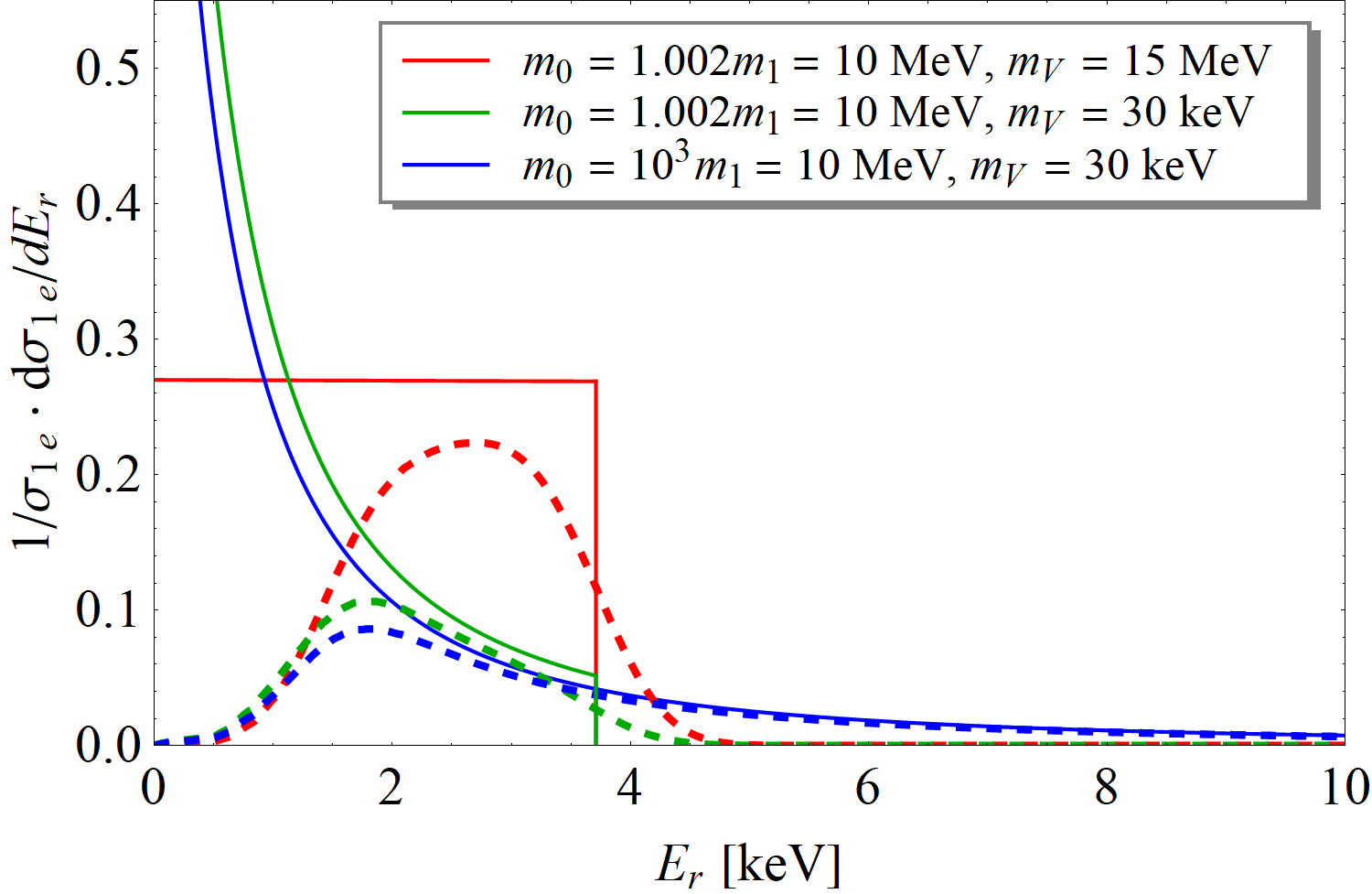}
    \caption{Unit-normalized electron recoil energy spectra (solid lines) in the VF case for three sets of mass values as shown in the legend.
    BDM and the mediator are a Dirac fermion and a massive vector.
    The dashed lines are the corresponding plots with detector resolution ($\sigma_{\rm res}=0.45$~keV) and efficiency reported in Ref.~\cite{Aprile:2020yad} and Ref.~\cite{Aprile:2020tmw}, respectively.}
    \label{fig:Espec}
\end{figure}

\noindent ($a$) \underline{Vector mediator}: 
We first consider the VF case (i.e., fermionic BDM), displaying example unit-normalized recoil energy spectra (solid lines) in figure~\ref{fig:Espec} with our benchmark parameter choices shown in the legend. 

In the first benchmark point (red), the BDM $\chi_1$ has a speed of $v_1=0.06c$, hence lies in the 68\% C.L.-favored region of Ref.~\cite{Kannike:2020agf} as also supported by the typical recoil energy of $\mathcal{O}({\rm keV})$.
Furthermore, since $m_1,m_V \gg m_e$ and $E_1 \approx m_1$, the spectral shape is almost flat over the allowed range in this limit:
\begin{equation}
    \frac{d\sigma_{1e}}{dE_r}\approx \frac{(g_\chi^Vg_e^V)^2 m_e m_1^2}{2\pi p_1^2m_V^4}\,, \label{eq:approx1}
\end{equation}
from which we find the total cross-section\footnote{Our expression has mass dependence different from the finding in Ref.~\cite{Fornal:2020npv}.
Ours is proportional to $m_e^2$ (vs. $m_e m_1$ in Ref.~\cite{Fornal:2020npv}), resulting in smaller estimates of cross-section.} to be
\begin{equation}
    \sigma_{1e}\approx \frac{(g_\chi^Vg_e^V)^2m_e^2}{\pi m_V^4}\,. \label{eq:xsform}
\end{equation}
This flat distribution can be distorted to a rising-and-falling shape by detector smearing and efficiency, as shown by the red dashed curve.  

For the second benchmark point (green), we choose a mediator $V$ lighter than electron.
Unlike the previous case, the expected recoil energy spectrum is rapidly falling off:
\begin{equation}
    \frac{d\sigma_{1e}}{dE_r} \approx \frac{(g_\chi^Vg_e^V)^2m_e m_1^2}{2\pi p_1^2(2m_eE_r+m_V^2)^2}\,, \label{eq:approx2}
\end{equation}
for which the total cross-section is dominated by the region of $E_r \to 0$.
The reason is because the differential cross-section in electron recoil momentum is peaking toward small $p_e(\ll m_e)$ due to the $t$-channel exchange of $V$ and this feature is more prominent for $m_V \ll m_e$~\cite{Kim:2020ipj}.
Once detector effects are included, events are expected to populate most densely around $2-3$~keV (see the green dashed curve).
However, a caveat to keep in mind is that too small $m_V$ values would lead most of events to lie below 2~keV since $d\sigma_{1e}/dE_r$ goes like $1/E_r^2$. 
Our numerical study suggests that $m_V \gtrsim 5$~keV would be favored by the data for the chosen $(m_0,m_1)$ pair. 

This observation motivates the third benchmark point (blue) where the BDM even lighter than electron acquires a significant boost factor.
An approximation similar to Eq.~\eqref{eq:approx2} goes through with $m_1^2$ replaced by $E_1^2$ since $E_1 \gg m_1$.
As also shown in figure~\ref{fig:Espec}, the differential spectrum is not much different from that of the second benchmark point, except a long tail beyond 7~keV which may not be appreciable at this earlier stage.
Moreover, the spectrum with detector effects (blue dashed) is quite similar to the second benchmark point.
This demonstrates that unlike the claim in Ref.~\cite{Kannike:2020agf} the favored region can be extended further below $\sim 0.1$~MeV and/or further beyond $v_1=0.3c$, as long as $m_V$ is smaller than $m_e$.
However, the preferred range of $m_V$ is more restricted than that in the second benchmark point.
Our numerical study shows that $m_V \gtrsim 50$~keV would result in more than half of events lying beyond 7~keV, so that $5 \lesssim m_V \lesssim 50$~keV would be favored for the chosen $(m_0,m_1)$ pair.

The cross-section $\sigma_{1e}$ also determines the mean free path $\bar{\ell}_1$ in the earth, which is given by $\sim 1/(\langle n_e \rangle \sigma_{1e})$ with $\langle n_e \rangle$ being the mean electron number density along the $\chi_1$ propagation line.
Here we assume that $\chi_1$ has negligible interactions with nuclei. 
If $g_e^V$ is too large (with $g_1^V$ set to be $\mathcal{O}(1)$), $\chi_1$ may scatter multiple times inside the earth before reaching the XENON1T detector located $\sim 1,600$~m underground, resulting in a substantial loss of energy that $\chi_1$ initially carries out.
The situation becomes worse if $\chi_1$ comes from the opposite side of the earth.
As shown in Eq.~\eqref{eq:Nsig}, $\mathcal{F}_1$ and $\sigma_{1e}$ are complementary to each other for a fixed $N_{\rm sig}$, i.e., a small $\mathcal{F}_1$ would be compensated by a large $\sigma_{1e}$ at the expense of multiple scattering of $\chi_1$.
This scenario was explored in Ref.~\cite{Fornal:2020npv}.
In our study, we rather focus on the opposite case where $\sigma_{1e}$ is small (hence no worries about the issue of too many scatterings) but sub-GeV (and smaller) $\chi_0$ allows a large flux of $\chi_1$.

\begin{figure}
    \centering
    \includegraphics[width=11cm]{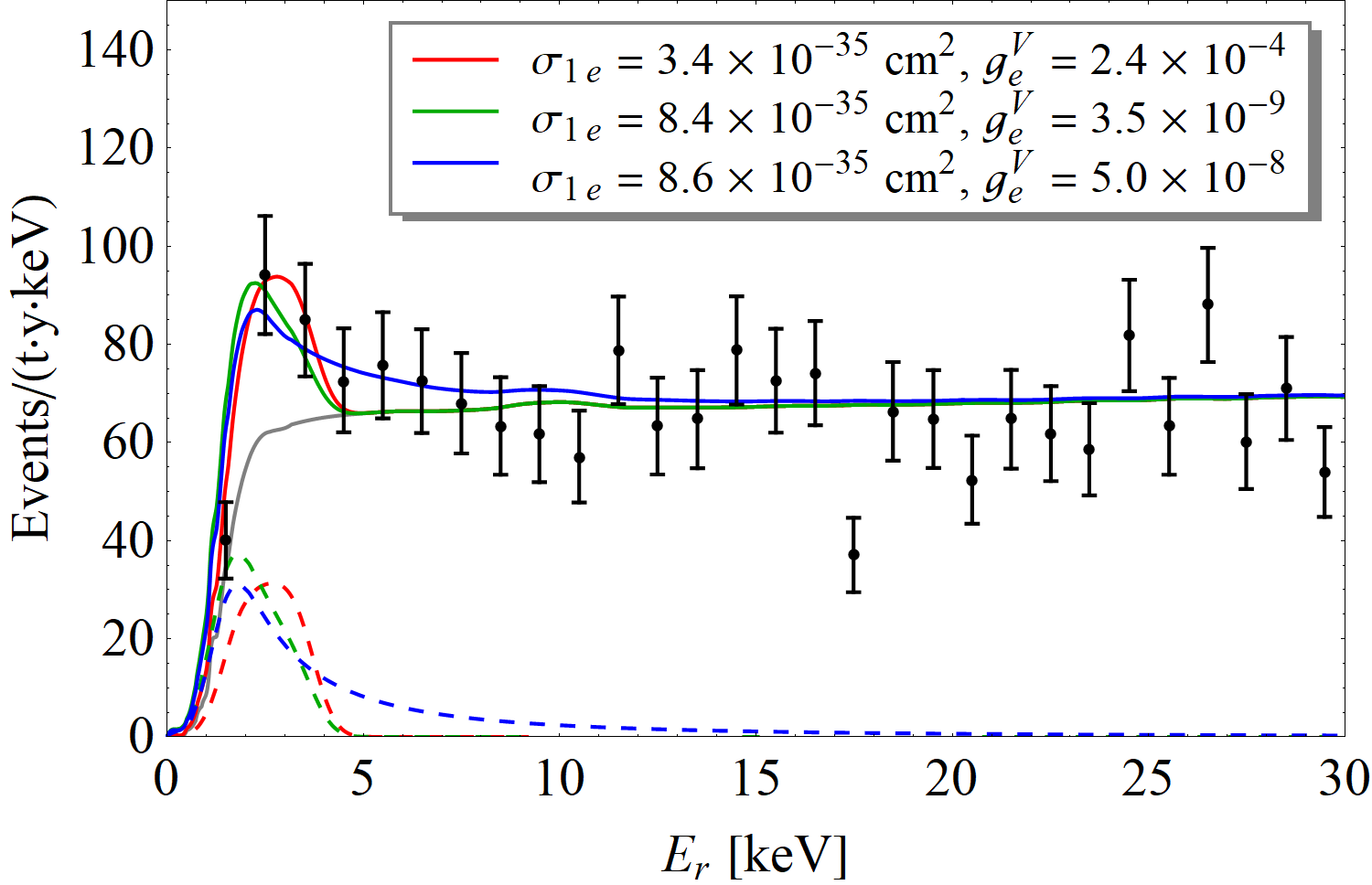}
    \caption{Sample energy spectra for the same benchmark mass spectra and particle spins (i.e., the VF case) as in figure~\ref{fig:Espec}.
    We assume $g_\chi^V=1$ and galactic BDM for which the flux is given by Eq.~\eqref{eq:flux}.
    The values of $\sigma_{1e}$ and $g_e^V$ associated with these fits are shown in the legend. The quoted $\sigma_{1e}$ are roughly consistent with the value of $\sigma_{1e}^{100}$ at $E_1=10$~MeV.
    The background model (with negligible tritium contributions) claimed by the XENON Collaboration and the data points are given by the gray line and the black dots, respectively.}
    \label{fig:fit}
\end{figure}

In figure~\ref{fig:fit}, we now show sample energy distributions for the three benchmark mass spectra taken in figure~\ref{fig:Espec}, assuming $g_\chi^V=1$ and galactic BDM whose flux is given by Eq.~\eqref{eq:flux}.
The values of $\sigma_{1e}$ and $g_e^V$ associated with these fits are shown in the legend.
The black dots and the gray line are the data points and the background model (with negligible tritium contributions) are imported from Ref.~\cite{Aprile:2020tmw}.

A few comments should be made for the quoted $\sigma_{1e}$ and $g_e^V$ values.
First, the required $\sigma_{1e}$ is of order $10^{-35}-10^{-34}~{\rm cm}^2$ resulting in more than ten thousand km ($\sim$ the diameter of the earth) of mean free path, i.e., at most a handful of $\chi_1$ scattering would arise inside the earth before reaching the XENON1T detector. See also the reference lines for $\sigma_{1e}^{100}$ and $\bar{\ell}_1$ in figure~\ref{fig:m0-m1}.
Second, there are mild differences among the quoted $\sigma_{1e}$ values although the BDM flux is fixed for all benchmark points.
As discussed earlier, the nominal scattering cross-section to explain the excess can be different due to the detector effects.
As suggested by figure~\ref{fig:Espec}, the green and blue curves are more affected by the detector efficiency since more events are expected to populate toward the lower energy regime.
Therefore, these two points typically demand a nominal BDM scattering cross-section greater than that for the other one.
Third, some of the reported $g_e^V$ values might be in tension with existing limits, depending on the underlying model details.
We will revisit this potential issue in the next section.

Finally, we briefly discuss how the variation in the dark matter spin affects the conclusions that we have made so far for the VF case.
We see that $\overline{|\mathcal{A}|}^2$ for the VS case is approximated to $16m_e^2E_1^2$ just like the VF case, and therefore expect similar spectral behaviors.
We find that the actual distributions look very similar to the corresponding ones with $\chi_1$ for the same mass choices, holding similar conclusions.

\medskip

\begin{figure}[t]
    \centering
    \includegraphics[width=0.496\linewidth]{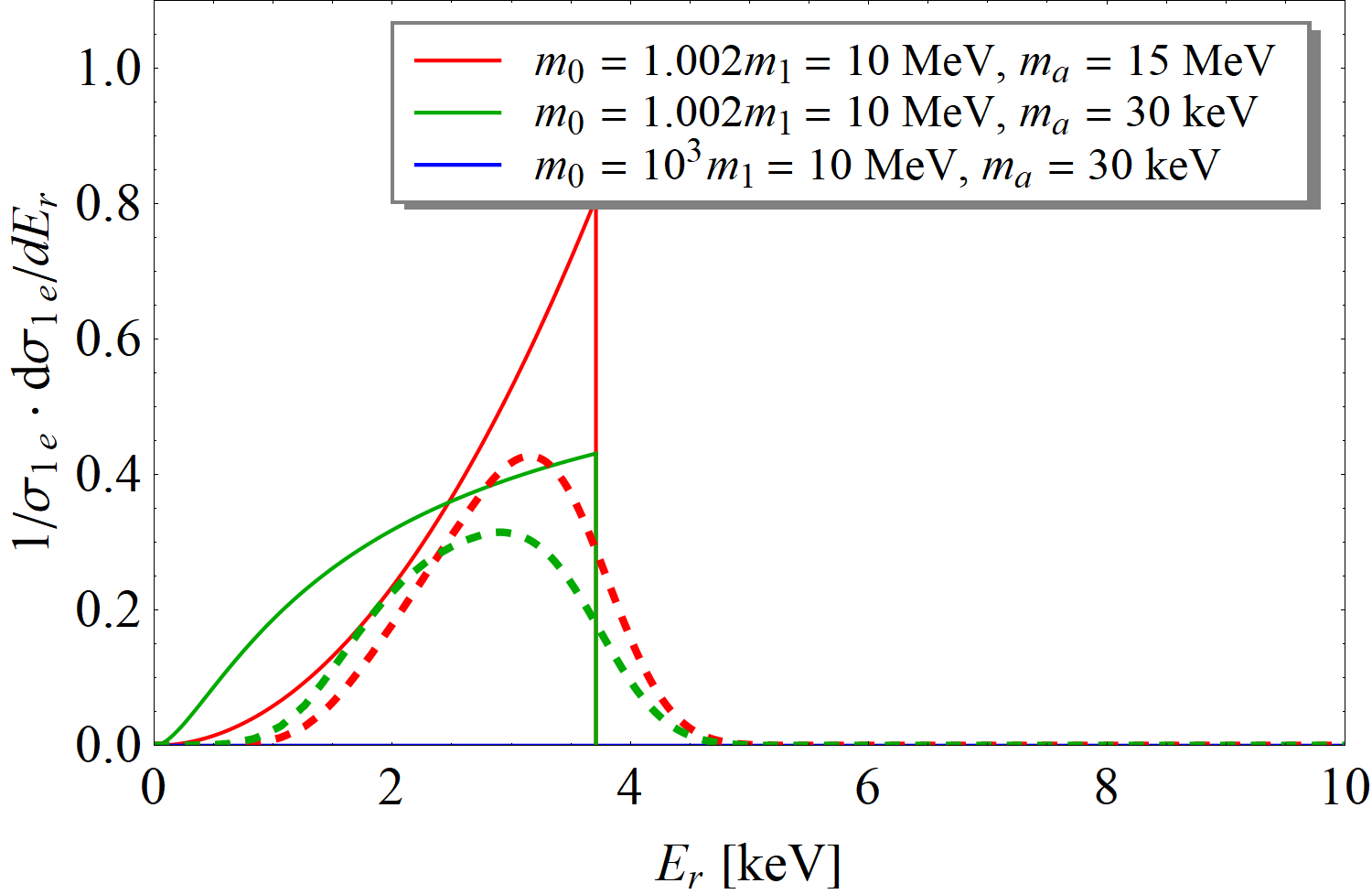} 
    \includegraphics[width=0.496\linewidth]{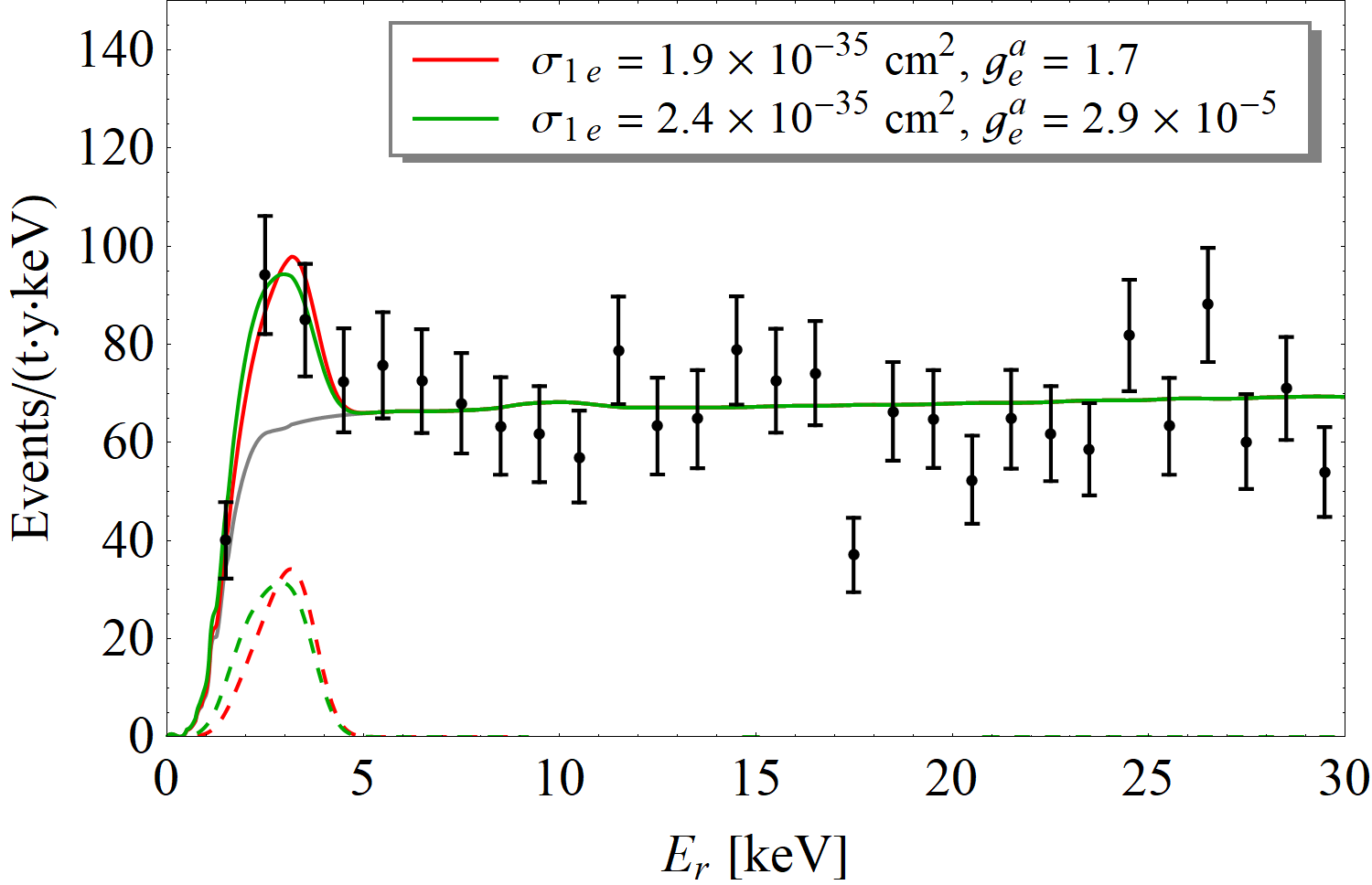}
    \caption{[Left] The corresponding unit-normalized plots with the same benchmark mass spectra as in figure~\ref{fig:Espec} but with fermionic BDM and pseudo-scalar mediator (i.e., the PF case).
    The solid and the dashed lines are the spectra without and with detector effects, respectively.
    For the ($iii$) region, the spectrum is rising very slowly toward $E_r^{\max}\approx 9.75$~MeV so that events with keV-scale recoil energy are very unlikely to arise and the corresponding blue curve appears invisible.
    [Right] Sample energy spectra for the first two benchmark mass spectra.
    We assume $g_\chi^a=1$ and galactic BDM for which the flux is given by Eq.~\eqref{eq:flux}.
    The values of $\sigma_{1e}$ and $g_e^a$ associated with these fits are shown in the legend.}
    \label{fig:Espec2}
\end{figure}

\noindent ($b$) \underline{Pseudo-scalar mediator}: 
We perform similar analyses for the three regions of mass space discussed in the previous section.
For fermionic dark matter $\chi_1$ (i.e., the PF case), we find that 
\begin{equation}
\hspace*{-0.1cm}    \frac{d\sigma_{1e}}{dE_r}\approx \left\{
    \begin{array}{ll}
    \dfrac{(g_\chi^a g_e^a)^2m_e E_r^2}{8\pi p_1^2 m_a^4}     & \hbox{ for }(i)  \\ [1.5em]
    \dfrac{(g_\chi^a g_e^a)^2m_e E_r^2}{8\pi p_1^2 (2m_e E_r+m_a^2)^2}     & \hbox{ for }(ii)\hbox{ and }(iii),
    \end{array}\right.
\end{equation}
and the corresponding energy spectra with the same benchmark mass spectra as in figure~\ref{fig:Espec} are shown in the left panel of figure~\ref{fig:Espec2}.
Unlike the vector mediator case, the differential cross-section rises in increasing $E_r$ due to the $E_r^2$ dependence in the numerators.
For ($i$) the recoil spectrum increases up to $E_r^{\max}$, whereas for ($ii$) it gradually saturates due to the competition with the $E_r$ dependence in the denominator.
All these expected behaviors are clearly shown by the solid red and the solid green curves in the left panel of figure~\ref{fig:Espec2}. 
Interestingly enough, the differential cross-section for ($ii$) becomes constant in the limit of $m_a\to 0$, and the $m_a$ dependence gets negligible.
Therefore, if a small $m_a$ is preferred by the data, it may be challenging to determine $m_a$.
For region ($iii$), exactly the same spectral behavior as in region ($ii$) is expected.
However, $E_r^{\max}$ approaches 9.75~MeV so that events accompanying keV-scale energy are very unlikely to arise.
Indeed, the blue curve clings to the $x$ axis.

This rising feature of the recoil energy spectra implies that less events are affected by the XENON1T detector efficiency unlike the ($ii$) and ($iii$) regions with a vector mediator.
In other words, nominal cross-sections differ not much from the corresponding fiducial cross-sections.
On the other hand, the total cross-section is much smaller than that of the vector mediator scenario for the same mass spectra and the same coupling strengths, because $E_r^2$ dependence (i.e., $\sim1-10~{\rm keV}^2$) is much smaller than $E_1^2$ dependence (see the discussions near Eqs.~\eqref{eq:approx1} and \eqref{eq:approx2}).
This implies that in order to obtain a required cross-section for a given BDM flux, a significantly larger coupling strength should be needed, compared to the corresponding value for the vector mediator.
The right panel of figure~\ref{fig:Espec2} shows sample energy spectra for the first two benchmark mass spectra with $g_\chi^a=1$ and galactic BDM, and clearly advocates all these expectations. The quoted $\sigma_{1e}$ are slightly smaller than the $\sigma_{1e}$ in figure~\ref{fig:fit}.
We also find that the required values of $g_e^a$ are larger than $g_e^V$ in figure~\ref{fig:fit} by roughly four orders of magnitude.
They may be strongly disfavored by the existing limits.
We again revisit this issue in the next section. 

When it comes to the case with scalar BDM (i.e., the PS case), we see that $E_r$ dependence in the numerator is linear so that the rising feature becomes mitigated.
In particular, for regions ($ii$) and $(iii$) the recoil energy distributions can be described by a rising-and-falling shape, so it is possible to find ranges of parameter space to explain the XENON1T excess.
We do not pursue an investigation to identify such parameter space here, reserving it for future work. 

\medskip

\noindent ($c$) \underline{Scalar mediator}: 
Given the discussions thus far, we are now equipped with enough intuitions to understand the scalar mediator case qualitatively.
In the SF case, $\overline{|\mathcal{A}|}^2$ behaves like $\sim m_e^2m_1^2$ for the ($i$) and ($ii$) regions, so the argument for the ($i$) and ($ii$) regions of the vector mediator scenario essentially gets through modulo numerical prefactors.
By contrast, the linear $E_r$ dependence can survive for the ($iii$) region, i.e., $\overline{|\mathcal{A}|}^2 \propto 2m_1^2+m_eE_r$, and as a result, the recoil energy spectrum can be of rising-and-falling shape like the ($ii$) and ($iii$) regions of the PS case.
In the SS case, $\overline{|\mathcal{A}|}^2\propto m_e^2m_1^2 = {\rm const.}$, so the overall expectations can be referred to those in the VF case except the fact that the scattering cross-sections are much smaller than those in the VF case for a given set of mass values and coupling strengths. 

\section{Discussions}

In this section, we discuss implications of our findings: fit parameter consistency with existing limits and scattering of BDM on xenon nuclei. 

As mentioned before, the quoted parameter values to explain the XENON1T excess may be in tension with existing bounds.
Identifying $V$ as a dark photon and considering the first benchmark point in figure~\ref{fig:fit}, we find that the $(m_V, g_e^V)$ pair is safe from the existing bounds.
In terms of the standard kinetic mixing parameter $\epsilon$, $g_e^V=2.4\times 10^{-4}$ is translated to $\epsilon=7.9\times10^{-4}$ which is not yet excluded by the latest limits~\cite{Banerjee:2019hmi}.
However, the parameter values for the other two benchmark points are strongly constrained by the limits from various astrophysical searches.
The same tension arises for the second benchmark point in the right panel of figure~\ref{fig:Espec2} with $a$ identified as, say axion-like particle.
Indeed, it was argued that there are ways to circumvent those astrophysical bounds that would rule out such dark photons and axion-like particles.
The main idea is that if the coupling constant and the mass parameter have effective dependence upon environmental conditions of astrophysical objects such as temperature and matter density, which are very different in the XENON1T experiment, the limits can be relaxed by several orders of magnitude~\cite{Jaeckel:2006xm,Ahlers:2006iz,Jaeckel:2010ni,An:2013yfc}.
There are several works to discuss relevant mechanisms in the context of specific particle physics models, e.g., Refs.~\cite{Khoury:2003aq, Masso:2005ym, Masso:2006gc, Mohapatra:2006pv, Dupays:2006dp, Brax:2007ak, Kim:2007wj}, for which concise summaries are referred to Refs.~\cite{Bonivento:2019sri, Dent:2019ueq}.
Furthermore, Ref.~\cite{An:2013yfc} pointed out that the energy loss process inside the stellar medium could be quenched because of absorption for large values of coupling.
Therefore, a careful check is needed to see if these parameter points are disfavored by the astrophysical bounds.
Finally, in regard to the $(m_a, g_e^a)$ values for the first benchmark point in the right panel of figure~\ref{fig:Espec2}, it seems that there are no existing searches that are sensitive to this parameter point to the best of our knowledge.
However, due to a relatively large size of coupling we expect that existing or near-future laboratory-based experiments such as accelerator experiments can test this parameter point.

Moving onto the second issue, one may ask whether BDM would scatter off a xenon nucleus and whether this dark matter interpretation would be contradictory to the null signal observation in the nuclear recoil channel at the XENON1T detector.
A possible solution is to assume that the mediator is ``baryo-phobic'' or ``electro-philic''.
Aside from model dynamics, we can check this issue using kinematics.
The maximum kinetic energy of a recoiling xenon nucleus $E_{r,{\rm Xe}}^{\max}$ is simply given by Eq.~\eqref{eq:maxE} with $m_e$ replaced by $m_{\rm Xe}$ and with $s$ approximated to $m_{\rm Xe}^2$.
For the first two benchmark mass points $p_1\approx 630$~keV gives $E_{r,{\rm Xe}}^{\max}\approx 6\times 10^{-3}$~keV, whereas for the last one $p_1=10$~MeV results in $E_{r,{\rm Xe}}^{\max}\approx 1.6$~keV.
Therefore, XENON1T is not sensitive enough to the dark matter signals from the three benchmark points in the nucleus scattering channel.
However, if $E_1$ increases, XENON1T starts to be sensitive to the signals belonging to region ($iii$) in the nucleus scattering channel, allowing for complementarity between the electron and nucleus recoil channels. 

\begin{figure}[t!]
    \centering
    \includegraphics[width=11cm]{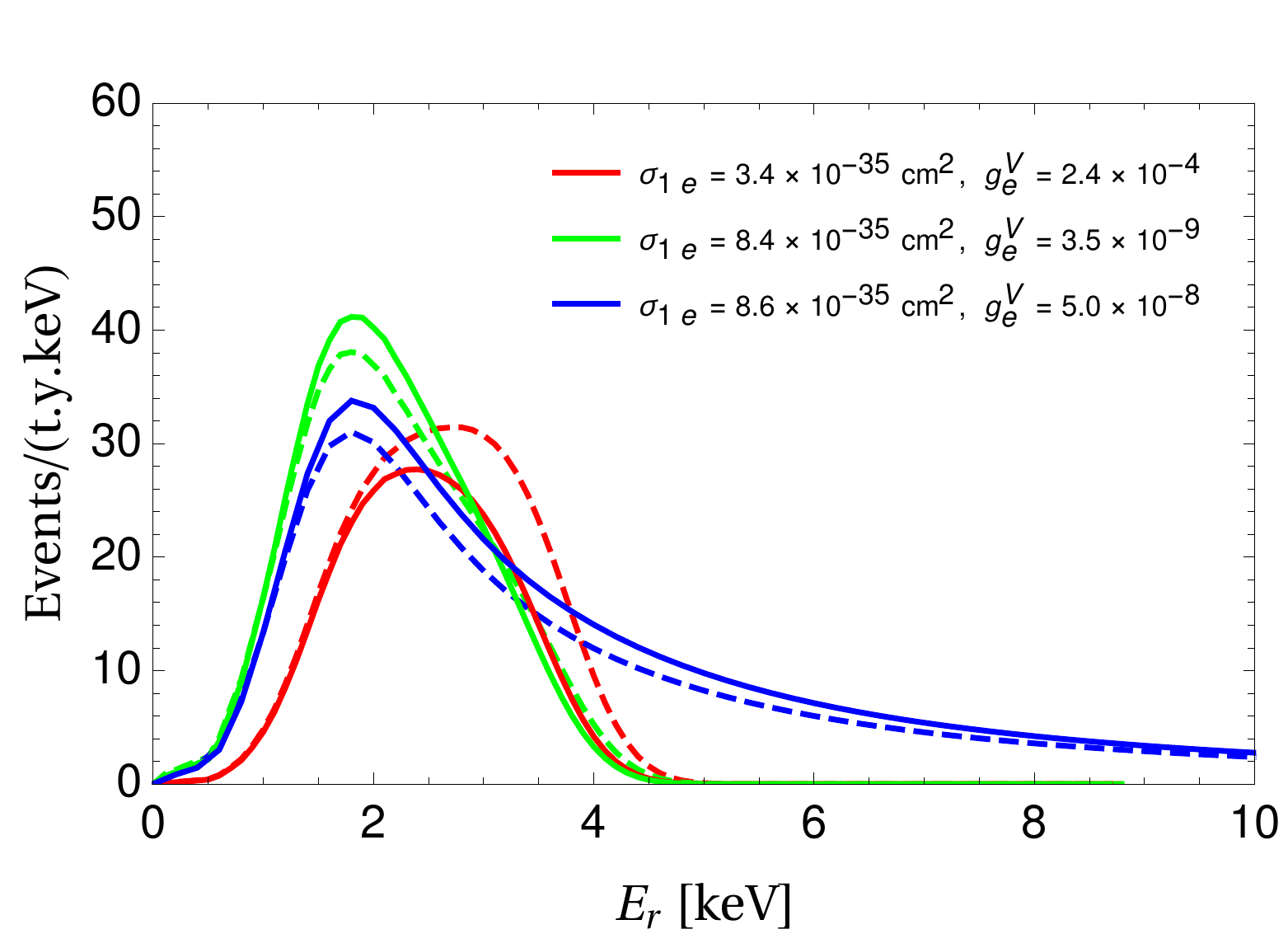}
    \caption{
    Sample energy spectra including the ionization factor for the same benchmark mass spectra and particle spins (VF case) as in figure~\ref{fig:Espec}. Three dashed curves are the same as those appearing in figure~\ref{fig:fit}, which already incorporate the detector resolution and the detector efficiency. The solid curves are the corresponding one further including effects of the ionization factor by the electrons in three outer shells. 
    }
    \label{fig:ionization}
\end{figure}

Finally, we would like to comment on the effects of the ionization form factor. In general, the form factors fall steeply with the momentum recoil, and therefore the ionization form factor strongly biases the scattering towards low-momentum recoil.
In addition, the form factor does not necessarily fall monotonically and thus could modify the recoil energy spectrum~\cite{Bunge:1993jsz, Kopp:2009et, Essig:2011nj, Lee:2015qva, Roberts:2016xfw, Catena:2019gfa, AKKMPS}.
The ionization factor can be calculated by using the Roothaan-Hatree-Fock wave function for the initial state electron \cite{Bunge:1993jsz} and applying the plane wave approximation for the final state electron.
We have followed the procedure described in Ref.~\cite{Cao:2020bwd,Kopp:2009et} to compute the ionization form factor for the interaction between BDM and the electrons in a xenon atom. We consider three outermost orbitals ($5p$, $5s$, and $4d$), with respective binding energies $\sim$12, 26 and 76 eV, which are known to be the dominant contribution \cite{Essig:2011nj,Lee:2015qva}.
As a cross-check, we have reproduced relevant results such as the ionization form factor from each shell and the differential recoil spectra for some physics examples as in Ref. \cite{Essig:2011nj,Lee:2015qva}.
We have also compared our approach against more sophisticated method where the final electron state is described by a positive energy continuum solution of the Schr\"odinger equation with a hydrogen potential~\cite{Roberts:2016xfw, Roberts:2019chv, Catena:2019gfa}.
We find that the plane wave approximation provides a reasonably good approximation for the low-momentum transfer as noted in Ref.~\cite{Roberts:2019chv, Catena:2019gfa}.  

In figure~\ref{fig:ionization} we show the energy spectra including the ionization factor for the same benchmark mass spectra and particle spins (VF case) as in figure~\ref{fig:Espec}.
Three dashed curves represent the energy spectra in figure~\ref{fig:fit}, which already incorporate the detector resolution and the detector efficiency, while the solid curves take into account effects of the ionization factor by considering 18 electrons in three outermost orbitals.
As the detector efficiency and resolution affect the shape of the energy spectra, the ionization form factor also gives additional distortion.
Nevertheless, the effects of the ionization factor in the shape of energy spectra are mild and the main features remain very similar. 
We find that the spectra for other particle spins (PF) also remain very similar to those in figure~\ref{fig:Espec2}.\footnote{More detailed analyses on the ionization form factor for the scattering of fast-moving dark matter and bound electrons are reserved for our future work~\cite{AKKMPS}.}

\section{Conclusions}
\begin{table}[t]
    \centering
    \begin{tabular}{c|c c c}
    \hline \hline
         &  Region ($i$)  & Region ($ii$) & Region ($iii$)\\
         \hline
        $\gamma_{\rm BDM}$ & $\approx 1$ &  $\approx 1$ & $\gg1$ \\
        \hline
        VF & \cmark (flat) & \comark (falling) & \comark (falling) \\
        VS & \cmark (flat) & \comark (falling) & \comark (falling) \\
        PF & \cmark (rising) & \cmark (rising) & \xmark (--) \\
        PS & \cmark (rising) & \comark (rising-and-falling) & \comark (rising-and-falling) \\
        SF & \cmark (flat) & \comark (falling) & \comark (rising-and-falling) \\
        SS & \cmark (flat) & \comark (falling) & \comark (falling) \\
        \hline \hline
    \end{tabular}
    \caption{A summary of our case studies.
    Conditions of each region are defined in Eq.~\eqref{eq:conditions}.
    $\gamma_{\rm BDM}$ denotes the Lorentz boost factor of BDM.
    \cmark and \comark ~indicate that one can find mass spectra to reproduce the XENON1T excess and satisfy the conditions of the associated regions, while for entries marked with \comark~a certain range of mediator mass may not reproduce the XENON1T excess.
    By contrast, \xmark~indicates that it is generally hard to find a mass spectrum to explain the excess.
    The general shape of expected recoil energy spectra is described in the parentheses.}
    \label{tab:summarytab}
\end{table}
The dark matter interpretation for the XENON1T anomaly is in favor of the existence of fast-moving or boosted dark matter component(s) in the present universe, which may require non-conventional dark matter dynamics.
We investigated various cases in which such dark matter of spin 1/2 and 0 interacts with electrons via the vector, pseudo-scalar, or scalar mediator in the context of the two-component boosted dark matter model as a concrete example.
Our findings are summarized in Table~\ref{tab:summarytab}.
We found that there exist a set of parameter choices to be compatible with existing bounds as well as to accommodate the anomaly.
In particular, the scales of mass and coupling parameters are sensitive to the mediator choice.
Our study further suggested that with appropriate choices of mediator and its mass, significantly boosted dark matter can be allowed on top of the moderately fast-moving dark matter.
Finally, we emphasize that the analysis method that we have proposed in this work is general, so we expect that it is readily applicable to the interpretation of observed data in other dark matter direct detection experiments. 

\medskip

\noindent {\bf \emph{Note Added. - }} We confirm that our total cross-section formula in Eq.~\eqref{eq:xsform} agrees with the corresponding expression in the updated version of Ref.~\cite{Fornal:2020npv}.

\medskip

\noindent {\bf \emph{Acknowledgement. - }} HA and KK acknowledge support from the US DOE, Office of Science under contract DE-SC0019474.
DK acknowledges support from DOE Grant DE-FG02-13ER41976/DE-SC0009913/DE-SC0010813. 
GM acknowledges support from DOE Grant Contract de-sc0012704. 
JCP acknowledges support from the National Research Foundation of Korea (NRF-2019R1C1C1005073 and NRF-2018R1A4A1025334). 
SS acknowledges support from the National Research Foundation of Korea (NRF-2020R1I1A3072747).

\bibliography{ref}{}
\bibliographystyle{JHEP} 

\end{document}